\documentclass [12pt]{article}
\usepackage {graphicx}
\begin{document}
\begin{center}
\textbf{PHYSICAL VACUUM PROPERTIES AND INTERNAL SPACE DIMENSION}
\end{center}

\bigskip

\begin{center}
\textbf{M.V. Gorbatenko, {\framebox{A.V. Pushkin}}}
\end{center}

\bigskip

\begin{center}
Russian Federal Nuclear Center - All-Russian Research Institute of 
Experimental Physics, Sarov, Nizhnii Novgorod region, 
\end{center}

\begin{center}
E-mail: {gorbatenko@vniief.ru}
\end{center}

\bigskip

\begin{center}
\textbf{Abstract}
\end{center}

\bigskip

The paper addresses matrix spaces, whose properties and dynamics are 
determined by Dirac matrices in Riemannian spaces of different dimension and 
signature. Among all Dirac matrix systems there are such ones, which 
nontrivial scalar, vector or other tensors cannot be made up from. These 
Dirac matrix systems are associated with the vacuum state of the matrix 
space. The simplest vacuum system realization can be ensured using the 
orthonormal basis in the internal matrix space. This vacuum system 
realization is not however unique. The case of 7-dimensional Riemannian 
space of signature 7(-) is considered in detail. In this case two basically 
different vacuum system realizations are possible: (\ref{eq1}) with using the 
orthonormal basis; (\ref{eq2}) with using the oblique-angled basis, whose base 
vectors coincide with the simple roots of algebra $E_{8} $. 

Considerations are presented, from which it follows that the least-dimen\-si\-on 
space bearing on physics is the Riemannian 11-dimensional space of signature 
$1\left( { -}  \right)\& 10\left( { +}  \right)$. The considerations consist 
in the condition of maximum vacuum energy density and vacuum fluctuation 
energy density.

\newpage

\bigskip

\section*{1. Introduction}

\bigskip

The problem under discussion in this paper consists in construction of 
matrix spaces (MS) in the Riemannian spaces of different dimensions and 
signatures and in the study of those MS properties, which can have a 
physical meaning as applied to the physical MS vacuum state. Before we 
formulate the problem more specifically, elucidate what is meant by the MS 
theory (see [1]-[3]). In so doing we restrict our consideration to minimum 
explanations needed for the consistent treatment.

The MS theory is a theory of internal degrees of freedom of Riemannian 
spaces. The properties of the internal degrees are introduced through Dirac 
matrices (DM) $\gamma _{A} $ satisfying relations
\begin{equation}
\label{eq1}
\gamma _{A} \gamma _{B} + \gamma _{B} \gamma _{A} = 2g_{AB} \cdot E.
\end{equation}
Here: $g_{AB} $ is the metric tensor of the Riemannian space of dimension $n$ 
($A,B = 1,2,...,n$); $\gamma _{A} ,\;\gamma _{B} $ are square matrices $N 
\times N$; $E$ is a unit matrix in the space of internal degrees of freedom.

The internal degrees of freedom are related, first, to transformations
\begin{equation}
\label{eq2}
\gamma _{A} \to {\gamma} '_{A} = S\left( {x} \right)\gamma _{A} S^{ - 
1}\left( {x} \right)
\end{equation}

\noindent
and, second, to the transition to the Riemannian spaces of higher dimensions 
and different signatures. 

Illustrate the MS properties by the example of the Riemannian space used in 
the general relativity. The space has dimension $n = 4$ and signature 
$1\left( { -}  \right)\& 3\left( { +}  \right)$. In this space, DM can be realized
above real, complex, quaternion and octonion fields. 

One of the best-known DM systems is that in the Majorana (real) 
representation:
\begin{equation}
\label{eq3}
\gamma _{0} = - i\rho _{2} \sigma _{1} ;\;\gamma _{1} = \rho _{1} ;\;\gamma 
_{2} = \rho _{2} \sigma _{2} ;\;\gamma _{3} = \rho _{3} .
\end{equation}
The DM realized above the complex field can be exemplified with: 

DM in the representation typically used in quantum electrodynamics:
\begin{equation}
\label{eq4}
\gamma _{0} = - i\rho _{3} ;\;\gamma _{1} = \rho _{2} \sigma _{1} ;\;\gamma 
_{2} = \rho _{2} \sigma _{2} ;\;\gamma _{3} = \rho _{2} \sigma _{3} .
\end{equation}

DM in the helicity representation:
\begin{equation}
\label{eq5}
\gamma _{0} = - i\rho _{1} ;\;\gamma _{1} = \rho _{2} \sigma _{1} ;\;\gamma 
_{2} = \rho _{2} \sigma _{2} ;\;\gamma _{3} = \rho _{2} \sigma _{3} .
\end{equation}

The well-known rho-sigma matrices $4 \times 4$ are used hereinafter in the 
notation of the expressions for DM. For the space used in the general 
relativity, the vector subscript $A$ will be denoted in a most habitual 
manner, viz. by lower case Greek letter ``$\alpha $'' ($\alpha = 0,1,2,3$). 
If an arbitrary field $\gamma _{\alpha}  \left( {x} \right)$ is taken, then 
scalar and tensor fields can be constructed from it, for example, the real 
scalar, such as

\begin{equation}
\label{eq6}
Sp\left( {\gamma ^{\alpha} \gamma _{\alpha} ^{ +} }  \right).
\end{equation}

Similar quantities will be, generally speaking, functions of coordinates. 
However, among the DM systems there are such systems, from which it proves 
impossible to construct coordinate dependent fields. Only trivial fields can 
be constructed from them, i.e. the fields, whose components reduce to 
integers constant in space. As it is easy to verify, such DM systems include 
systems (\ref{eq3}), (\ref{eq4}), (\ref{eq5}). 
We will term these DM systems as vacuum DM (VDM). We 
will therewith use this term not only in the case of the Riemannian space 
used in the general relativity, but also in any other case. 

The vacuum DM systems in spaces of different dimensions and signatures are 
just the subject of consideration in this paper. It will be shown that in 
most cases the VDM are realized through using the orthonormal basis in the 
internal space. However, for some dimensions and signatures this vacuum 
system realization is not unique. The case of 7-dimensional Riemannian space 
of signature 7(-) is considered in detail. In this case two basically 
different vacuum system realizations are possible: (\ref{eq1}) with using the 
orthonormal basis; (\ref{eq2}) with using the oblique-angled basis, whose  
base vectors coincide with the simple roots of algebra $E_{8} $. In the second 
case the vacuum DM system offers a unique feature. The feature is that it 
has maximum vacuum energy density and vacuum fluctuation energy density in 
comparison with other vacuum DM systems. The compactification of the 
internal 7-dimensional space leads to the 11-dimensional Riemannian space of 
signature $1\left( { -}  \right)\& 10\left( { +}  \right)$. 

The question arises: Can any other Riemannian space of a dimension higher 
than 11 exist, which would admit a denser ``zero'' energy packing than the 
11-dimensional space of signature $1\left( { -}  \right)\& 10\left( { +}  
\right)$ with compactified 7-dimensional subspace? From the general theory 
of lattices it is known that densest packings are admitted by 8-dimensional 
lattice $E_{8} $ and 24-dimensional Leech lattice. Hence, the obtained 
result regarding the dimensional regularization admits, in principle, only 
one correction. The correction can appear only in consideration of the 
lattice VDM configurations described by the Leech lattice. However, even on 
the transition to a space, in which the VDM are described by the simple 
roots of the Leech lattice, the 11-dimensional space of signature $1\left( { 
-}  \right)\& 10\left( { +}  \right)$ will preserve the sense of the space, 
in which the local minimum of action on vacuum is realized. 

The obtained results have a direct bearing, for example, on a so-called 
problem of dimensional regularization in multidimensional schemes 
(supergravity; superstring theories). In 
essence, this is a forcible argument in favor of the fact that the dimension 
of the World is 11 and its signature is of form $1\left( { -}  \right)\& 
10\left( { +}  \right)$.

\section*{2. MS theory}

\bigskip

In what follows the following facts pertaining to the MS theory are 
considered known\footnote{ The facts mentioned are substantiated in refs. 
[1]-[3].} :

The realization of DM above the complex and quaternion field reduces to the 
special cases of the realizations above the real field, but in Riemannian 
spaces of a higher dimension. So in what follows it is sufficient to 
restrict our consideration to the realization of DM only above the real 
field. 

In the realization of DM above the real field, without loss of generality we 
can consider only the Riemannian spaces of odd dimension $n$, 
\begin{equation}
\label{eq7}
n = 2k + 1,
\end{equation}

\noindent
where $k$ is a positive integer. The signature of the Riemannian spaces 
under consideration should therewith be of form 
$k(-)\&(k+1)(+)$ or differ from it by a 
number of ``minuses'' multiple of four.

The DM which can be introduced in the Riemannian space possessing the above 
mentioned properties are square matrices $N \times N$, with the $N$ being 
related with $k$ as
\begin{equation}
\label{eq8}
N = 2^{k}.
\end{equation}
The hermitizing (or anti-hermitizing) matrix $D$ determined as
\begin{equation}
\label{eq9}
D\gamma _{\alpha}  D^{ - 1} = - \gamma _{\alpha} ^{ +}  ,
\end{equation}

\noindent
can be introduced in any MS. The matrix $D$ is a Hermitean matrix and, 
hence, it can be used as a metric in the internal space, that is can be used 
to introduce the Hermitean quadratic form.

The best-known vacuum DM type is the systems of DM possessing the following 
properties (we will refer to them as standard VDM):

- They can be written in the form of monomials in terms of the rho-sigma 
matrices.

- They possess certain symmetric, real properties.

- They have either zeroes or numbers $ \pm 1$ as their elements.

\section*{3. Physical criteria for comparison of VDM properties}

\bigskip

Ask ourselves the question: Are there any criteria to select those 
multidimensional Riemannian spaces with different values of $n$ and all 
possible signatures among all the ones, in which the VDM could be comparable 
to the physical vacuum?

The answer to the question is of the fundamental nature, as it bears on the 
problem of dimensional regularization in the multidimensional theories. 
Without an answer to this question the multidimensional theories become a 
subject of investigation, whose relation to the reality is unknown.

It is clear that the question posed cannot be answered without going beyond 
the standard VDM class. In fact, when solving the question of physical 
advantages or disadvantages of the VDM, we can appeal either to any scalar 
characteristics of the internal space like 
\begin{equation}
\label{eq10}
\frac{{1}}{{N}}Sp\left( {D} \right)
\end{equation}

\noindent
or to any integrals over the elementary cell, which the VDM is constructed on. The 
quantity of the second type can be exemplified with
\begin{equation}
\label{eq11}
\frac{{1}}{{NV_{Cell}} }\int\limits_{Cell} {Z\sqrt {det\left( {D} \right)} 
\;dZ} .
\end{equation}
The integration in (\ref{eq11}) is performed over all the elementary cell points in the 
internal space. 

The quantities of type (\ref{eq10}) will be interpreted as the vacuum energy density 
and those of type (\ref{eq11}) as the vacuum fluctuation energy density.

For all standard VDM these characteristics are the same,
\begin{equation}
\label{eq12}
\frac{{1}}{{N}}Sp\left( {D} \right) = 1,
\end{equation}
\begin{equation}
\label{eq13}
\frac{{1}}{{NV_{Cell}} }\int\limits_{Cell} {Z\sqrt {det\left( {D} \right)} 
\;dZ} = Const.
\end{equation}

\section*{4. Explicit form of nonstandard VDM system} 

\bigskip

We have found that the standard VDM do not constitute a unique VDM class. It 
turns out that for some dimensions and signatures basically another type of 
the VDM systems can exist, which we will term as an alternative VDM type. As 
this paper reveals, the comparison between the standard and alternative VDM 
properties allows us to find those criteria, which lead to essentially 
unique solution to the dimensional regularization problem.

The analysis of the DM structure in spaces of different dimensions and 
signatures, which is performed by us in ref. [3], suggests that the 
nonstandard VDM type can exist only with a definite signature of the matrix 
degree of freedom subspace part, which is described by the nonstandard  
DM type. From the general theory of lattices suitable for physical use (see, 
e.g., [4]) it is known that the lattice generated in the matrix degree of 
freedom space should be autodual. It is also known (see, e.g., [5]) that the 
least dimension of the Riemannian space admitting introduction not only of a 
dual lattice based on the orthonormal basis, but also a basically different 
lattice, is 7. The ``different'' lattice structure is therewith based on the 
simple root vectors of the algebra $E_{8} $. 

The satisfaction of all the above conditions needed for the nonstandard VDM 
appearance cannot be achieved with arbitrary Riemannian space dimension and 
signature. We have found that the 7-dimensional definite space admitting the 
description with DM $8 \times 8$ can be included as a compactified internal 
space in the ordinary 4-dimensional space of signature $1\left( { -}  
\right)\& 3\left( { +}  \right)$. The signature of the 7-dimensional 
subspace should be $7\left( { -}  \right)$ and that of the complete space 
$1\left( { -}  \right)\& 10\left( { +}  \right)$. 

The simplest method to prove the existence of the alternative VDM along with 
the standard VDM system is to present the explicit form of both the systems. 
We will do this now.

Below are the explicit forms of two DM systems in the 7-dimensional space of 
signature $7\left( { -}  \right)$. Either system is a set of seven matrices 
$8 \times 8$. The first DM system refers to the standard VDM category; we 
will denote it by $\{ \gamma_{A} \}  \quad \left( {A = 1,2, \cdots ,7} 
\right)$. The second system refers to the alternative VDM category; we will 
denote it by $\left\{ {\bf g}_{A}  \right\} \quad \left( {A = 1,2, \cdots ,7} 
\right)$.

The system $\{ \gamma _{A} \} $ can be written in terms of sigma matrix $2 
\times 2$ and rho-sigma matrix $4 \times 4$, i.e. as
\begin{equation}
\label{eq14}
\left. {\begin{array}{l}
 {\gamma_{1} = \sigma_{3} \otimes i\rho_{2} \sigma_{1} \quad \quad 
\gamma_{2} = \sigma_{3} \otimes i\sigma_{2} \quad \quad \gamma_{3} = 
\sigma_{3} \otimes i\rho _{2} \sigma_{3}}  \\ 
 {\gamma_{4} = \sigma_{1} \otimes i\rho_{1} \sigma_{2} \quad \quad 
\gamma_{5} = \sigma_{1} \otimes i\rho_{2} \quad \quad \gamma_{6} = 
\sigma_{1} \otimes i\rho_{3} \sigma_{2}}  \\    
 {\gamma_{7} = i\sigma _{2} \otimes E\quad \quad D = E \otimes 
E} \\ 
 \end{array}}  \right\}.
\end{equation}
    
Recall that the anti-hermitizing matrix $D$ is determined as 
\begin{equation}
\label{eq15}
\gamma _{A}^{ +}  = - D\gamma _{A} D^{ - 1}.
\end{equation}

Now this is the explicit form of DM $\left\{ {\bf g}_{A}  \right\}$.
$$
{\bf g}_1=\begin{array}{|c|c|c|c|c|c|c|c|}\hline
&&-1&&&&&\\ \hline
-1&-1&-1&-2&-3&-2&-1&-2\\ \hline
1&&&&&&&\\ \hline
-1&-1&-1&-1&-1&-1&-1& \\ \hline
&&&&&&1&\\ \hline
1&2&3&4&5&3&1&2\\ \hline
&&&&-1&&&\\ \hline
&&-1&-2&-2&-1&-1&-1\\ \hline
\end{array}
$$
$$
{\bf g}_2=\begin{array}{|c|c|c|c|c|c|c|c|}\hline
&&1&&&&&\\ \hline
&-1&-2&-2&-3&-2&-1&-2\\ \hline
-1&&&&&&&\\ \hline
&-1&-1&-1&-1&-1&& \\ \hline
&&&&&&-1&\\ \hline
1&2&3&4&4&3&2&2\\ \hline
&&&&1&&&\\ \hline
&&-1&-2&-2&-1&&-1\\ \hline
\end{array}
$$
$$
{\bf g}_3=\begin{array}{|c|c|c|c|c|c|c|c|}\hline
-1&-2&-2&-2&-3&-2&-1&-2\\ \hline
1&1&1&&&&&\\ \hline
&&1&2&3&2&1&2\\ \hline
-1&-1&-2&-3&-4&-3&-1&-2 \\ \hline
1&2&3&4&5&4&2&2\\ \hline
&&&&-1&-1&-1&\\ \hline
-1&-2&-3&-4&-4&-2&-1&-2\\ \hline
&-1&-2&-2&-2&-2&-1&-1\\ \hline
\end{array}
$$
$$
{\bf g}_4=\begin{array}{|c|c|c|c|c|c|c|c|}\hline
-1&-2&-3&-4&-5&-4&-2&-2\\ \hline
1&2&3&4&5&3&1&2\\ \hline
-1&-2&-3&-4&-4&-2&-1&-2 \\ \hline
1&2&3&3&3&2&1&2\\ \hline
-1&-2&-2&-2&-3&-2&-1&-2\\ \hline
1&1&1&2&3&2&1&2\\ \hline
&&-1&-2&-3&-2&-1&-2\\ \hline
&1&1&1&2&1&1&1\\ \hline
\end{array}
$$
$$
{\bf g}_5=\begin{array}{|c|c|c|c|c|c|c|c|}\hline
&&&&&&1&\\ \hline
-1&-2&-3&-4&-4&-3&-2&-2\\ \hline
&&&&-1&&&\\ \hline
1&2&2&3&4&2&1&2 \\ \hline
&&1&&&&&\\ \hline
&-1&-2&-2&-3&-2&-1&-2\\ \hline
-1&&&&&&&\\ \hline
&&&1&2&2&1&1\\ \hline
\end{array}
$$
$$
{\bf g}_6=\begin{array}{|c|c|c|c|c|c|c|c|}\hline
-1&-2&-3&-4&-4&-2&-1&-2\\ \hline
1&2&3&4&4&3&2&2\\ \hline
-1&-2&-3&-4&-5&-4&-2&-2\\ \hline
1&1&2&3&4&3&1&2 \\ \hline
&&-1&-2&-3&-2&-1&-2\\ \hline
&1&2&2&3&2&1&2\\ \hline
-1&-2&-2&-2&-3&-2&-1&-2\\ \hline
&&&1&2&1&1&1\\ \hline
\end{array}
$$
\begin{equation}
\label{eq16}
{\bf g}_7=\begin{array}{|c|c|c|c|c|c|c|c|}\hline
&&&&1&&&\\ \hline
&&&&&1&&\\ \hline
&&&&&&1&\\ \hline
&-1&-2&-3&-4&-3&-2&-2 \\ \hline
-1&&&&&&&\\ \hline
&-1&&&&&&\\ \hline
&&-1&&&&&\\ \hline
2&3&4&5&6&4&2&3\\ \hline
\end{array}
\end{equation}

Systems $\{ \gamma _{A} \} $, $\left\{ {\bf g}_{A}  \right\}$ satisfy 
relations (\ref{eq1}), which in this case have form
\begin{equation}
\label{eq17}
\gamma _{A} \gamma _{B} + \gamma _{B} \gamma _{A} = - 2\delta _{AB} ;\quad 
{\bf g}_{A} {\bf g}_{B} + {\bf g}_{B} {\bf g}_{A} = - 2\delta _{AB} 
\end{equation}

\noindent
and, hence, according to the Pauli theorem, are related as
\begin{equation}
\label{eq18}
\gamma _{A} \to {\bf g}_{A} = R\gamma _{A} R^{ - 1}.
\end{equation}

The matrix $R$ is found uniquely (with accuracy to common multiplier) and 
takes the form

\begin{equation}
\label{eq19}
R=\begin{array}{|c|c|c|c|c|c|c|c|}\hline
1&-1&&&&&&\\ \hline
&1&-1&&&&&\\ \hline
&&1&-1&&&&\\ \hline
&&&1&-1&&& \\ \hline
&&&&1&-1&&\\ \hline
&&&&&1&-1&\\ \hline
&&&&&&1&-1\\ \hline
-1/2&-1/2&-1/2&-1/2&-1/2&1/2&1/2&1/2\\ \hline
\end{array}
\end{equation}
\begin{equation}
\label{eq20}
R^{-1}={{1}\over{2}}\begin{array}{|c|c|c|c|c|c|c|c|}\hline
1&0&-1&-2&-3&-2&-1&-2\\ \hline
-1&0&-1&-2&-3&-2&-1&-2\\ \hline
-1&-2&-1&-2&-3&-2&-1&-2\\ \hline
-1&-2&-3&-2&-3&-2&-1&-2 \\ \hline
-1&-2&-3&-4&-3&-2&-1&-2\\ \hline
-1&-2&-3&-4&-5&-2&-1&-2\\ \hline
-1&-2&-3&-4&-5&-4&-1&-2\\ \hline
-1&-2&-3&-4&-5&-4&-3&-2\\ \hline
\end{array}
\end{equation}

For DM $\left\{ {\bf g}_{A}  \right\}$ the anti-hermitizing matrix $\bf D$ is 
found from relation
\begin{equation}
\label{eq21}
{\bf D} = R^{ - 1}{}^{ +} DR^{ - 1}
\end{equation}

\noindent
and proves equal to
\begin{equation}
\label{eq22}
{\bf D}=\begin{array}{|c|c|c|c|c|c|c|c|}\hline
2&3&4&5&6&4&2&3\\ \hline
3&6&8&10&12&8&4&6\\ \hline
4&8&12&15&18&12&6&9\\ \hline
5&10&15&20&24&16&8&12 \\ \hline
6&12&18&24&30&20&10&15\\ \hline
4&8&12&16&20&14&7&10\\ \hline
2&4&6&8&10&7&4&5\\ \hline
3&6&9&12&15&10&5&8\\ \hline
\end{array}
\end{equation}

The inverse matrix is
\begin{equation}
\label{eq23}
{\bf D}^{-1}=\begin{array}{|c|c|c|c|c|c|c|c|}\hline
2&-1&&&&&&\\ \hline
-1&2&-1&&&&&\\ \hline
&-1&2&-1&&&&\\ \hline
&&-1&2&-1&&& \\ \hline
&&&-1&2&-1&&-1\\ \hline
&&&&-1&2&-1&\\ \hline
&&&&&-1&2&\\ \hline
&&&&-1&&&2\\ \hline
\end{array}
\end{equation}

The question arises: In what way can the 7-dimensional subspace, 
whose vacuum state is described by latticed VDM ${\bf g}_A$, be integrated 
with the ordinary 4-dimensional space-time used in the general relativity? 
In principle, there are two possibilities. The first possibility is to make the 
3-dimensional general relativity subspace a subspace of the 7-dimensional space. 
This possibility leads to the new question: Why are only 3 dimensions of 7 observable? 
We do not know any answer to the question, so we will not consider this possibility. 
The second possibility is to assume that the complete space is 11-dimensional 
and factorized to the direct product of two subspaces: 4-dimensional general 
relativity subspace and 7-dimensional subspace. It is this possibility 
that is considered by us in what follows.

From refs. [1]-[3] it follows that in the MS of dimension $n = 2k + 1$ the 
real DM are of dimension $N \times N$, where $N = 2^{k}$. Among the set of 
all MS there is a subset of such MS, in which the DM structure is of the 
following form:
\begin{equation}
\label{eq24}
Matrix\left( {N \times N} \right) \Rightarrow Matrix\left( {2^{k - 2} \times 
2^{k - 2}} \right) \otimes Matrix\left( {2^{2} \times 2^{2}} \right),
\end{equation}

As applied to the 11-dimensional MS of signature $1\left( { -}  \right)\& 
10\left( { +}  \right)$ the DM systems have dimension $32 \times 32$ and 
subset (\ref{eq24}) is factorized as follows:
\begin{equation}
\label{eq25}
Matrix_{32 \times 32} = Matrix_{8 \times 8} \otimes Matrix_{4 \times 4} .
\end{equation}

Here is the explicit form of the DM LS: 
\begin{equation}
\label{eq26}
\begin{array}{l}
 \Gamma _{\alpha}  = {\bf e} \otimes \gamma _{\alpha}  \quad \quad 
\alpha = 0,1,2,3 \\ 
 \Gamma _{4} = {\bf g}_{1} \otimes \gamma \quad \quad \Gamma _{5} = {\bf g}_{2} 
\otimes \gamma \quad \quad \quad \quad \Gamma _{6} = {\bf g}_{3} 
\otimes \gamma \quad \quad \quad \quad \quad \quad \quad \quad \quad \\ 
 \Gamma _{7} = {\bf g}_{4} \otimes \gamma \quad \quad \quad \quad \quad 
\quad \Gamma _{8} = {\bf g}_{5} \otimes \gamma \quad \quad \Gamma _{9} = 
{\bf g}_{6} \otimes \gamma \quad \\ 
 \Gamma _{10} = {\bf g}_{7} \otimes \gamma \quad \\ 
 \end{array}
\end{equation}
Here $\gamma $ denotes matrix $4 \times 4 \quad \gamma \equiv \gamma _{0} \gamma 
_{1} \gamma _{2} \gamma _{3} $ (it is typically denoted as $\gamma_{5} $), 
${\bf e}$ is the unit matrix $8 \times 8$, and ${\bf g}_{1} 
,...,{\bf g}_{7} $ are matrices $8 \times 8$ determined by relations (\ref{eq16}). 
The ${\bf g}_{1} ,...,{\bf g}_{7} $ are the DM for the 7-dimensional space 
of signature $7\left( { -}  \right)$. When the 7-dimensional space as a 
subspace is included in the 11-dimensional space, the signature of the 
7-dimensional space changes from $7\left( { -}  \right)$ to $7\left( { +}  
\right)$. 

The table below lists the characteristics of the matrix space being 
discussed in this paper.

\vskip7mm

Table 1. List of characteristics of the MS under discussion in this paper
\vskip7mm
\noindent
\begin{tabular}{|p{7cm}|p{5cm}|}\hline
Riemannian space dimension& 
$11$ \\ \hline
Riemannian space signature& 
$ 1\left( { -}  \right)\& 10\left( { +}  \right)$  \\ \hline
Matrix degree of freedom space dimension& 
$32 \times 32 $ \\ \hline
Dimension of the compactified part of the Riemannian subspace& 
$7$ \\ \hline
Signature of the compactified part of the Riemannian subspace& 
$7\left( { +}  \right) $ \\ \hline
Matrix structure in the matrix degree of freedom space& 
$M_{32 \times 32} = M_{8 \times 8} \otimes M_{4 \times 4} $ \\ \hline
\end{tabular}

\section*{5. Comparison of different VDM properties}

\bigskip

Compare properties of systems $\left\{ {\gamma _{A}}  \right\}$, $\left\{ 
{\bf g}_{A}  \right\}$.

- First and foremost note that in either system the DM elements are integer 
numbers. Matrices $D$, $D^{ - 1}$,$\bf D$, ${\bf D}^{ - 1}$ are integer matrices as 
well.

- Matrix $R$ is not a unitary matrix, therefore the system $\left\{ {\bf g}_{A}  
\right\}$ definitely does not reduce to $\left\{ {\gamma _{A}}  
\right\}$ through the orthogonal transformation either of base vectors of 
the 7-dimensional space or those of the internal 8-dimensional space. 

- The structure of DM $\left\{ {\gamma _{A}}  \right\}$ and matrices $D$, 
$D^{ - 1}$ possesses the following properties:

- Each of the above matrices is written in the form of a monomial, which is 
a direct product of the rho-sigma matrices.

- As few as one element is nonzero in each row and each column, with the 
nonzero element being equal to $ \pm 1$. 

- Each of the above matrices is either symmetrical or antisymmetrical. 

\underline {The necessary condition for conversion of DM to VDM}:

For some DM system to be a vacuum system, it is necessary that both matrix 
$D$ and matrix $D^{ - 1}$ be integer matrices.

Statement.

If matrices $D$, $D^{ - 1}$ are integer, then their determinants are the 
same and equal either to +1 or -1. 

The statement is proved as follows. First the evident fact is noted: in an 
integer matrix its determinant is also an integer. Let
\begin{equation}
\label{eq27}
det\left( {D^{ - 1}} \right) \equiv p,
\quad
det\left( {D} \right) \equiv q,
\end{equation}

\noindent
where $p,\;q$ are integers. As $D$, $D^{ - 1}$ are matrices reciprocal to 
each other, then the following relation should hold: 
$det\left( {D^{ - 1}} \right) = \frac{{1}}{{det\left( {D} \right)}},$ that is 
$pq = 1$. 

The only combinations of integers $p,\;q$ satisfying the relation are
\[
p = q = \pm 1.
\]
Thus, for VDM 
\begin{equation}
\label{eq28}
det\left( {D^{ - 1}} \right) = det\left( {D} \right) = \pm 1.
\end{equation}

The following statements constitute the consequence of the above VDM 
definition:

1. If some DM system has been found to be a VDM, then the system derived 
from the original through arbitrary $T$- and $L$-transformations will also 
be a VDM.

2. There is a $T$-representation, in which the VDM is constant over the 
entire space. 

3. From VDM it is impossible to construct not only a nontrivial tensor, but 
also a nontrivial tensor function. 

The straightforward check of properties of the matrices $\left\{ {\gamma 
_{A}}  \right\}$, $\left\{ {\bf g}_{A}  \right\}$ indicates that the 
systems $\left\{ {\gamma _{A}}  \right\}$, $\left\{ {\bf g}_{A}  \right\}$ 
satisfy the above definition of VDM. At the same time, these two VDM systems 
are not interrelated with the $L$- and $O$-transformations, with accuracy to 
which the VDM are determined.

\section*{6. Weyl group}

\bigskip

From the general theory of lattices it is known (see, e.g., [5]) that the 
densest packings are admitted by the 8-dimensional lattice $E_{8} $ and 
24-dimensional Leech lattice $\Lambda _{24} $. Hence, the case of the 
8-dimensional lattice $E_{8} $ definitely has an independent meaning. The 
results can be used later on in consideration of a more complex case of the 
24-dimensional Leech lattice $\Lambda _{24} $.

The basis of the lattice $E_{8} $ consists of eight simple roots in this 
algebra. It is common practice to depict the simple root sets in the form of 
Dynkin's diagrams (see, e.g., [6]-[8]). In Fig. 1 the diagrams are shown for 
two root systems: $e^{k}$ for the orthonormal algebra; $r^{a}$ for the 
algebra $E_{8} $. 

\begin{figure}[htbp]
\centering
\includegraphics*[bbllx=0.17in,bblly=0.17in,bburx=7.60in,bbury=2.68in,scale=0.7]{fig.bmp}
\end{figure}

\bigskip

Fig. 1. Dynkin's diagrams for the basic systems $e^{k}$, $r^{a}$

\bigskip

Each basic system consists of 8 vectors denoted with the open circle in the 
diagrams. The dark circle is correspondent with the vector norm equal to 1 
and the open circle with that equal to $\sqrt {2} $. If there is no line 
between the circles, then they are orthogonal. If the circles are connected 
with a solid line, then the angle between the respective vectors is 
$120^{0}$.

The complete set of the root vectors of the algebra $E_{8} $ consists of 240 
vectors and any system of the simple roots in the algebra of 8 vectors. From 
the set of all roots the simple root system is constructed nonuniquely. By 
way of example consider matrix $\tilde {R}$ describing a simple root system 
distinct from that described by matrix $R$ of form (\ref{eq19}).

\begin{equation}
\label{eq29}
\tilde {R} = \begin{array}{|c|c|c|c|c|c|c|c|}\hline
1&-1&&&&&&\\ \hline
&1&-1&&&&&\\ \hline
&&1&-1&&&&\\ \hline
&&&1&-1&&& \\ \hline
&&&&1&-1&&\\ \hline
&&&&&1&-1&\\ \hline
1/2&1/2&1/2&1/2&1/2&-1/2&1/2&1/2\\ \hline
&&&&&1&1&\\ \hline
\end{array}
\end{equation}

Matrix $W$relating matrices $R$ of form (\ref{eq19}) and $\tilde {R}$ of form (\ref{eq29}) 
according to relation

\begin{equation}
\label{eq30}
\tilde {R} = W \cdot R,
\end{equation}

\noindent
is of the form:

\begin{equation}
\label{eq31}
W = = \begin{array}{|c|c|c|c|c|c|c|c|}\hline
-1&&&&&&&\\ \hline
&-1&&&&&&\\ \hline
&&-1&&&&&\\ \hline
&&&-1&&&& \\ \hline
1&2&3&4&4&2&1&2\\ \hline
&&&&&1&&\\ \hline
-1&-2&-3&-4&-5&-4&-2&-3\\ \hline
-1&-2&-3&-4&-5&-3&-1&-2\\ \hline
\end{array}
\end{equation}

The above matrix $W$ is one of the elements of a so-called Weyl group for 
the root system in algebra $E_{8} $. Any matrix $W$ is distinguished with 
two properties: the determinant equal to one and integer elements. None of 
the matrices $W$ is orthogonal. The set of the matrices $W$ determining the 
transition from one simple root system to another comprises a finite group 
of unimodular matrices above the integer field. The order of the Weyl group 
coincides with the quantity of different simple root systems and is equal to 
(see, e.g., [6])

\begin{equation}
\label{eq32}
\rho = 2^{14} \cdot 3^{5} \cdot 5^{2} \cdot 7 = 696\;729\;600.
\end{equation}

\section*{7. Comparison of VDM by magnitudes of physical quantities for vacuum}

\bigskip

Thus, we have obtained two solution types for VDM. One type corresponds to 
the choice of Euclidean basis in the local internal spaces ${\rm 
M}_{8}^{local} $, the second type corresponds to the choice of Euclidean 
basis from the simple roots in algebra $E_{8} $. It is natural to pose the 
question as to what of these VDM describes the physical vacuum.

One of the criteria for the selection could be the magnitude of the simplest 
scalar

\begin{equation}
\label{eq33}
Sp\left( {D^{ - 1}} \right) = 8;\quad Sp\left( {\bf D}^{ - 1} \right) = 16
\end{equation}

\noindent
calculated for both the VDM. For VDM the magnitudes of scalars (\ref{eq33}) are the 
same, which is easy to find, as the values of $Sp\left( {M} \right)$, where 
$M$ is the polarization density matrix. Its average value $Sp\left( {M} 
\right)$ in the MS vacuum state has the meaning of background (``zero'') 
physical vacuum energy density $E^{vac}$,

\begin{equation}
\label{eq34}
Sp\left( {M} \right) = E^{vac}.
\end{equation}

From this it follows that the ``zero'' vacuum energy packing density 
$E^{vac}$ is two times higher for the VDM constructed on the roots of the 
algebra $E_{8} $ than that for the VDM constructed on the orthonormal basis. 
A denser packing is more beneficial energetically, hence, the vacuum 
solution is that VDM, in which the matrix $D^{ - 1}$ is of form (\ref{eq23}).

Another possible criterion for the selection of the true vacuum solution for 
VDM can be elementary cell averaged magnitudes of scalars

\begin{equation}
\label{eq35}
\tilde {E}^{vac\;fluc} \equiv \left\langle {Sp\left( {Z^{ +} DZ} \right)} 
\right\rangle _{cell} ;
\quad
\tilde {\bf E}^{vac\;fluc} \equiv \left\langle {Sp\left( {Z^{+} {\bf D}Z} \right)} 
\right\rangle _{cell} .
\end{equation}

Averaging $\left\langle {} \right\rangle _{cell} $ in the evaluation of (\ref{eq35}) 
implies that not only the values of the matrix $Z$ at the lattice nodes, but 
also those at all points within the elementary cell are considered. Therefore, 
whereas scalars (\ref{eq33}) characterize the background vacuum energy density, 
scalars (\ref{eq35}), apparently, have the meaning of vacuum fluctuation energy 
density. 

In space ${\rm M}_{8}^{local} $ single out the elementary cells corresponding to 
the orthonormal lattice and lattice $E_{8} $. For the orthonormal lattice 
the cell volume is
\begin{equation}
\label{eq36}
V = \int\limits_{0}^{1} {dx^{1}} \cdots \int\limits_{0}^{1} {dx^{8}} = 1.
\end{equation}

For the lattice $E_{8} $ the cell volume is
\begin{equation}
\label{eq37}
{\bf V} = \int\limits_{0}^{\sqrt {2}}  {dy^{1}} \cdots \int\limits_{0}^{\sqrt {2} 
} {dy^{8}} \sqrt {det(\bf D)} = 2^{4}.
\end{equation}

In (\ref{eq37}) the integration is performed in the limits from $0$ to $\sqrt {2} $, 
since the length of each of the basic spinor matrices is $\sqrt {2} $ (see 
Fig. 1); besides, $\sqrt {det(\bf D)} = 1$ is taken into account.

Now calculate values (\ref{eq37}) for the orthonormal cell and for cell $E_{8} $.

The value of the quadratic form $\tilde {E}^{vac\;fluc} $ over the elementary 
orthonormal cell is
\begin{equation}
\label{eq38}
\tilde {E}^{vac\;fluc} = \left( {{1}\over{V}} 
\right)\;\;\int\limits_{0}^{1} {dx^{1}} \cdots \int\limits_{0}^{1} {dx^{8}} 
\left\{ {{x^{1}}^{2} + {x^{2}}^{2} + ... +{x^{8}}^{2}} \right\} = 
\frac{{8}}{{3}}.
\end{equation}

The value of the quadratic form $\tilde {\bf E}^{vac\;fluc}$ over the elementary cell 
constructed on the simple roots in the algebra $E_{8} $ is
\begin{equation}
\label{eq39}
\begin{array}{l}
 \tilde {{\bf E}}^{vac\;fluc} = \left( {\frac{{1}}{{\rm V}}} 
\right)\int\limits_{0}^{\sqrt {2}}  {dy^{1}} \cdots \int\limits_{0}^{\sqrt 
{2}}  {dy^{8}} \left\{ 2{y^{1}}^{2} + 2{y^{2}}^{2} + 2{y^{3}}^{2} + 2{y^{4}}^{2}\right. +\\ 
\left. +2{y^{5}}^{2}+ 2{y^{6}}^{2} + 2{y^{7}}^{2} + 
 + 2y^{8}{}^{2} - 2y^{1}y^{2} - 2y^{2}y^{3} - 2y^{3}y^{4}\right. -\\ 
\left. -2y^{4}y^{5} - 2y^{5}y^{6} - 2y^{5}y^{8} - 2y^{6}y^{7} \right\} = 
{{11}\over{3}} \\ 
 \end{array}
\end{equation}

The comparison of (\ref{eq39}) with (\ref{eq38}) shows that the density of elementary 
cell space 
filling with vacuum fluctuation energy is ${{11} \mathord{\left/ {\vphantom 
{{11} {8}}} \right. \kern-\nulldelimiterspace} {8}}$ times higher for the 
lattice $E_{8} $ than that for the orthonormal lattice, that is

\begin{equation}
\label{eq40}
{{\tilde {\bf E}^{vac\;fluc}} \mathord{\left/ {\vphantom {{\tilde 
{E}^{vac\;fluc}} {\tilde {E}^{vac\;fluc}}}} \right. 
\kern-\nulldelimiterspace} {\tilde {E}^{vac\;fluc}}} = {{11} \mathord{\left/ 
{\vphantom {{11} {8}}} \right. \kern-\nulldelimiterspace} {8}}
\end{equation}

Note that this result agrees with the results of the consideration of the 
classic problem of the search for a method of densest sphere packing in 
8-dimensional space (see, e.g., [5]). 

\section*{8. Conclusion}

\bigskip

The paper has demonstrated that a Riemann space, in which it becomes 
possible to introduce a nonstandard type of autodual vacuum DM system, is the 
space of dimension 7 and signature $7\left( { -}  \right)$. For the 
Riemannian space to be integrated with the conventional Riemann space of the 
general relativity, it should be assumed that the 7-dimensional space is not 
only an internal subspace of a complete space, but also a compactified 
space. The resultant complete Riemannian space is of dimension 11 and 
signature $1\left( { -}  \right)\& 10\left( { +}  \right)$.

In the 7-dimensional space of the constructed 11-dimensional space, the DM 
vacuum system is the system $\left\{ {\bf g}_{A}  \right\} \quad \left( {A = 
1,2, \cdots ,7} \right)$, whose explicit form is given in (\ref{eq16}). The DM 
$\left\{ {\bf g}_{A}  \right\}$ are matrices $8 \times 8$ and have integer 
elements. In the 8-dimensional space of the matrix degrees of freedom of DM 
$\left\{ {\bf g}_{A}  \right\}$ the simple root vectors of the algebra 
$E_{8} $ are used for the basis. The transition of the orthonormal basis to 
he basis of the simple root vectors of the algebra $E_{8} $ is performed by 
matrix $R$ of form (\ref{eq19}).

The vacuum state corresponding to the DM based on the simple root vectors of 
the algebra $E_{8} $ is in fact degenerate, with the degeneracy degree being 
the same as the order of Weyl group, a group of symmetry of the root system 
of the algebra. For $E_{8} $ the Weyl group order is determined as (\ref{eq32}). 

The introduction of the lattice VDM based on the simple root vectors of the 
algebra $E_{8} $ changes the MS vacuum state structure in a qualitative 
sense. It turns out that ``zero'' energy density for these VDM is higher 
than that for the VDM in their conventional representation. From the general 
principles of physics it follows that the vacuum state of the highest 
``zero'' energy density is most stable and, hence, it is this state that 
must be used for the description of physical effects. This, in its turn, 
singles out the 11-dimensional space of signature $1\left( { -}  \right)\& 
10\left( { +}  \right)$ among other spaces of lower dimensions and other 
signatures. 

Thus, this paper has arrived at a direct answer to the question of what 
space dimensions and signatures should be considered 
in the physical theories, where the need arises to refer to 
multidimensional Riemannian spaces (supergravity; superstring theories). 
A candidate for the highest-priority 
consideration in the multidimensional version of the interaction theories 
should be the 11-dimensional Riemannian space of signature $1\left( { -}  
\right)\& 10\left( { +}  \right)$ as a space of a most stable vacuum state. 
The stability of the 
vacuum state excludes appearance of tachyons causing a reconstruction of the state.
The group of symmetry of the internal 7-dimensional subspace is the group 
$O_{8} $ well-known in the superstring theory (see, e.g., [4]). 

The results of this paper on the Dirac matrices constructed 
on the basis from lattice $E_{8} $ can be also used in the construction 
of Dirac matrices on the basis from Leech lattice $\Lambda_{24}$. 
This follows from the following two facts. First, according 
to [9], [10], the Leech lattice admits splitting 16+8, 
i.e. the splitting into a sublattice of two $E_{8} $ and a sublattice 
of one $E_{8} $. Second, the internal Riemannian space dimension 
can be in principle not necessarily equal to 7, 
it can be multiple of 7. With dimension 21, the internal subspace 
description will require the Dirac matrices constructed from matrices 
${\bf g}_A$ in the form of their direct sum.

Pay attention to the fact that to obtain the above nontrivial results it was 
not necessary to resort to exotic schemes, to complication of the class of 
numbers, groups of symmetry, geometries, etc. It was sufficient to use the 
adequately studied lattice properties, the most fundamental class of 
numbers, i.e. integer real numbers, and intuitively clear matrix space 
vacuum state properties, i.e. impossibility to produce any coordinate 
dependent tensor quantities from the Dirac matrix vacuum systems. After that 
it turned out that the requirements to the DM vacuum systems can be met not 
only on the way of using the standard orthonormal basis, but also the 
oblique-angled basis corresponding to the system of simple roots of the 
algebra $E_{8} $. The properties of lattices, integers, and vacuum state 
that have been used in this paper are of ``super-quantum'' nature in the 
sense that they will hardly be measured in any quantization procedure. 

The work was carried out under partial financial support by the 
International Science and Technology Center (ISTC Project KR-677).

\bigskip

\section*{References}

\bigskip

[1] M.V. Gorbatenko, A.V. Pushkin. \textit{VANT; Ser.: Teor. i Prikl. Fiz}. 
\textbf{1}, 49 (1984).

\noindent
[2] M.V. Gorbatenko. \textit{TMF}. \textbf{103}, \textbf{1}, 32 (1995).

\noindent
[3] M.V. Gorbatenko, A.V. Pushkin. \textit{VANT; Ser.: Teor. i Prikl. Fiz}. 
\textbf{2-3}, 61 (2000).

\noindent
[4] M. Green, J. Schwartz, E. Witten. \textit{Superstring Theory. Introduction.} 
Cambridge University Press (1987). 

\noindent
[5] N.J.A. Sloane. \textit{Scientific American}. Vol. {\bf 250}, No. 1 (1984).

\noindent
[6] D.P. Zhelobenko. Compact Lie groups and their representations. Moscow, 
Nauka Publishers (1970).

\noindent
[7] R.E. Behrends, J. Dreitlein, C. Fronsdal, W. Lee. \textit{Rev. Mod. 
Phys.,} {\bf 34}, No. 1, 1 (January 1962). 

\noindent
[8] L.S. Pontryagin. \textit{Continuous groups.} Moscow, Gostekhizdat 
Publishers (1954).

\noindent
[9] G. Chapline. Phys. Letters. Vol. 158B, No 5, 393 (1985).

\noindent
[10] J. Lepowsky and A. Meurman, J. Algebra 77 (1982) 484.

\end{document}